\documentclass{midl} 

\title[POWDR]{POWDR: Pathology-preserving Outpainting with Wavelet Diffusion for 3D MRI}

\midlauthor{\Name{Fei Tan\nametag{$^{1}$}} \orcid{0000-0003-2423-631X} \Email{Fei.Tan@gehealthcare.com}\\
\addr $^{1}$ GE HealthCare, San Ramon, CA, USA \AND
\Name{Ashok Vardhan {Addala}\nametag{$^{1}$}} \Email{AshokVardhan.Addala@gehealthcare.com}\\
\Name{Bruno Astuto Arouche {Nunes}\nametag{$^{1}$}} \orcid{0000-0002-8890-7963} \Email{Bruno.Nunes@gehealthcare.com}\\
\Name{Xucheng Zhu\nametag{$^{2}$}} \orcid{0000-0002-5537-5542} \Email{Xucheng.Zhu@gehealthcare.com}\\
\addr $^{2}$ GE HealthCare, Menlo Park, CA, USA \AND
\Name{Ravi Soni\nametag{$^{1}$}} \Email{Ravi.Soni@gehealthcare.com}\\
}

\begin{document}

\maketitle

\begin{abstract}
Medical imaging datasets often suffer from class imbalance and limited availability of pathology-rich cases, which constrains the performance of machine learning models for segmentation, classification, and vision–language tasks. To address this challenge, we propose POWDR, a pathology-preserving outpainting framework for 3D MRI based on a conditioned wavelet diffusion model. Unlike conventional augmentation or unconditional synthesis, POWDR retains real pathological regions while generating anatomically plausible surrounding tissue, enabling diversity without fabricating lesions.

Our approach leverages wavelet-domain conditioning to enhance high-frequency detail and mitigate blurring common in latent diffusion models. We introduce a random connected mask training strategy to overcome conditioning-induced collapse and improve diversity outside the lesion. POWDR is evaluated on brain MRI using BraTS datasets and extended to knee MRI to demonstrate tissue-agnostic applicability. Quantitative metrics (FID, SSIM, LPIPS) confirm image realism, while diversity analysis shows significant improvement with random-mask training (cosine similarity reduced from 0.9947 to 0.9580; KL divergence increased from 0.00026 to 0.01494). Clinically relevant assessments reveal gains in tumor segmentation performance using nnU-Net, with Dice scores improving from 0.6992 to 0.7137 when adding 50 synthetic cases. Tissue volume analysis indicates no significant differences for CSF and GM compared to real images.

These findings highlight POWDR as a practical solution for addressing data scarcity and class imbalance in medical imaging. The method is extensible to multiple anatomies and offers a controllable framework for generating diverse, pathology-preserving synthetic data to support robust model development.

\end{abstract}


\section{Introduction}


The development of medical imaging analysis methodologies using deep learning is hindered by two major challenges: the scarcity of pathological data and pronounced class imbalance. Acquiring large-scale, diverse datasets is resource-intensive, and many pathological conditions occur at low prevalence, limiting the availability of representative samples. These constraints impede the training of robust and generalizable models. Consequently, developing strategies to augment the availability of pathology-rich images without incurring additional acquisition costs is critical.

Generative models have emerged as a promising solution to bridge this gap. Among them, diffusion models have demonstrated strong capabilities in synthesizing realistic medical images \cite{khader_medical_2023}. While unconditional diffusion models can generate plausible images, they often fail to capture rare pathological patterns \cite{dar_unconditional_2025}. Existing approaches such as inpainting have primarily focused on reconstructing healthy regions \cite{kofler_brain_2024} or performing super-resolution across slices \cite{kang_deep_2021}. In contrast, outpainting—the process of expanding an image beyond its original boundaries while preserving a given region—remains underexplored in medical imaging. Outpainting offers a unique advantage: it allows retention of real pathological regions while generating surrounding healthy tissue, thereby creating diverse and clinically relevant training samples. This work leverages conditional diffusion models to achieve this goal.

Latent Diffusion Models (LDMs) \cite{rombach_high-resolution_2022} have become the standard for image synthesis; however, they struggle to recover high-frequency details \cite{lozupone_latent_2025, falck_fourier_2025}, often producing blurry outputs. Wavelet-based diffusion models \cite{gerdes_gud_2024, friedrich_wdm_2025} address this limitation by decomposing images into high- and low-frequency components and conditioning generation on these channels, thereby enhancing fine structural details. Although wavelet diffusion has been applied to contrast synthesis, its potential for outpainting tasks in medical imaging remains untapped.

In this study, we introduce POWDR, Pathology-preserving Outpainting with conditioned Wavelet Diffusion for 3D MRI. The main contributions of our work are threefold:
\begin{enumerate}
  \item We propose a novel outpainting framework that conditions on pathological regions and synthesizes surrounding anatomy, enabling generation of full MRI volumes with real pathology.
  \item We demonstrate the effectiveness of POWDR through a comprehensive evaluation using quantitative image quality metrics (FID, SSIM, LPIPS) and clinically relevant assessments, including segmentation accuracy and tissue volume comparisons against real patient data.
  \item We demonstrate that our approach is tissue-agnostic, validating its applicability on brain and knee MRI and discussing its extensibility to other anatomical regions.
\end{enumerate}

\section{Methods}

\subsection{Network Design}

\subsubsection{Training}

Our architecture (\figureref{fig:fig1}) builds upon the conditional wavelet diffusion framework \cite{friedrich_cwdm_2024} with several enhancements for 3D medical image synthesis. In the forward diffusion process, the original 3D MRI volume $y_0 \in \mathbb{R}^{H\times W \times D} $ is decomposed using a digital wavelet transform (DWT) with Haar kernel \cite{stankovic_haar_2003} into $x_0 \in \mathbb{R}^{8\times \frac{H}{2} \times \frac{W}{2} \times \frac{D}{2}}$ which has eight subbands:
$$x_0=DWT(y_0 )=\{LLL,LLH,LHL,LHH,HLL,HLH,HHL,HHH\}$$

representing low- and high-frequency components across different orientations. Gaussian noise $\epsilon \sim \mathcal{N}(0,\,I)\ $ is progressively added according to a linear $\beta$-schedule ranging from 0.0001 to 0.02. Resulting in the noised image $x_t=\sqrt{(\bar{\alpha}_t )x_0+(1-\bar{\alpha}_t )} \epsilon$, where $\bar{\alpha}_t=\prod_{i=1}^t\alpha_i$  , $\alpha_t=1-\beta_t$. 

For conditional denoising, we employ a 3D ResUNet \cite{jha_resunet_2019} architecture. The conditioning image is also transformed into wavelet subbands and concatenated with the noisy target subbands $[\widetilde{x}_t, c]$ at each denoising step, forming 16 input channels. The conditioning image during training can be either a masked pathology region or a randomly masked region, as detailed in the Dataset section. The network outputs $\widetilde{x}_{t-1}$ and estimates $\widetilde{x}_0$, containing 8 channels, and the mean squared error (MSE) in the wavelet domain is adopted as loss function
$$L = MSE_{wavelet} = \lVert x_0  - \widetilde{x}_0 \rVert ^2$$

Other key training parameters include, dropout rate = 0.1, training 200,000 iterations, diffusion time step = 1000, ResUNet base channel size = 64, channel multipliers = [1, 2, 2, 4, 4], AdamW optimizer \cite{loshchilov_decoupled_2019}, learning rate = $1\times 10^{-5}$. All experiments were conducted on an NVIDIA H100 GPU (80 GB VRAM).

\begin{figure}[htbp]
\floatconts
  {fig:fig1}
  {\caption{Architecture of POWDR: a pathology-preserving outpainting framework using conditioned wavelet diffusion model for 3D MRI. The pipeline begins with wavelet decomposition and Gaussian noise addition during forward diffusion. In the conditional denoising step, the wavelet-transformed masked pathology is concatenated with the noisy input and processed by an attention-based ResUNet. The inverse wavelet transform recombines frequency components to reconstruct the synthesized image}}
  {\includegraphics[width=0.9\linewidth]{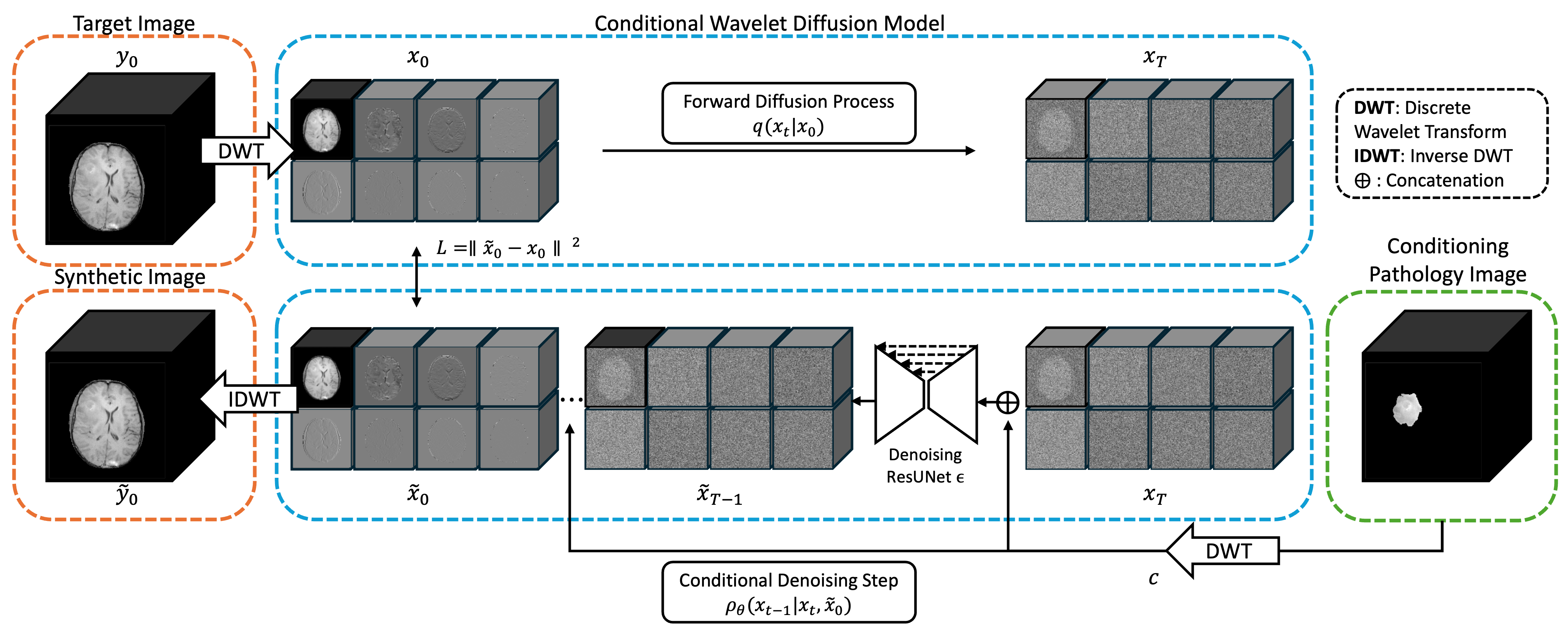}}
\end{figure}

\subsubsection{Sampling}
During inference, masked pathology region conditioning is achieved by element-wise multiplication of the pathology mask with the corresponding intensity volume (e.g., tumor mask with brain image or cartilage mask with knee image). This approach ensures that pathological regions are preserved while enabling synthesis of surrounding anatomy. During inference, we apply 1,000 denoising steps following the Denoising Diffusion Probabilistic Model (DDPM) sampling strategy \cite{ho_denoising_2020}. The generated sample, initially in latent space, is then reconstructed into the image domain using the Inverse Discrete Wavelet Transform (IDWT).
$$\widetilde{y}_0 = IDWT(\widetilde{x}_0)$$

\subsection{Datasets} 

\subsubsection{Brain MRI}

We use the BraTS 2023 dataset for training (1251 cases, T1n contrast) and BraTS 2024 additional cases for evaluation (100 cases for sampling, 171 cases for segmentation testing) \cite{karargyris_federated_2023, baid_rsna-asnr-miccai_2021, bakas_advancing_2017, menze_multimodal_2015}. The same approach can be expanded to other contrasts (T2w, T1 contrast-enhanced, T2-FLAIR). 

The default training strategy utilizes masked tumor region as training condition. To increase conditioning diversity, we introduce additional training strategy, which generates random connected masks multiplied by image volume as training conditions. These masks are 6-connected and sampled based on tumor volume distribution from the training set to match realistic tumor sizes. All images are intensity-normalized to the 1st-99th percentile, and cropped into $224 \times 224 \times 155$ matrix size and 1mm isotropic resolution. During sampling for both training scenarios, only the tumor region is used for conditioning to ensure pathology preservation. 

\subsubsection{Knee MRI}

We further validate our approach on a knee MRI dataset comprising 768 3D volumes collected across multiple scanners and two distinct MRI sequences. The first subset is a public dataset acquired using Double Echo Steady State (DESS) sequences on Siemens scanners, originally described by \cite{balamoody_comparison_2010} and used in a segmentation challenge \cite{desai_international_2021}. This subset includes 176 MRI volumes from 88 patients, with manual 3D segmentation masks for four tissue compartments—patellar, tibial, and femoral cartilage, as well as the menisci—generated by a single expert reader \cite{desai_international_2021}.

The second subset is an internal dataset of 592 studies from 275 unique patients. These volumes were acquired using CUBE, a high-resolution 3D Fast Spin Echo (FSE) sequence developed by GE Healthcare, and previously utilized by \cite{astuto_automatic_2021} for anomaly and object detection. Scans were obtained on five 3-T MRI scanners (GE Healthcare, Waukesha, WI) between 2006 and 2018, spanning multiple scanner models and sites.

All knee MRI volumes were preprocessed to ensure consistency: cropped to a common field-of-view ($144 \times 144 \times 80$ mm), resized to $288 \times 288 \times 160$ voxels, and resampled to 0.5 mm isotropic resolution. Cartilage and meniscus masks were combined and dilated by 3 pixels using a spherical kernel to account for thin structures, and element-wise multiplied with image volume as condition. Both healthy and pathological cases were included in training, while sampling was conditioned on meniscus tear pathology. This setup demonstrates that the model can leverage both healthy and pathological images during training while enabling pathology-conditioned synthesis—a capability particularly valuable when pathological cases are scarce.

Approval for data use was obtained from the Osteoarthritis Initiative (OAI) (https://nda.nih.gov/oai) for the DESS dataset and from the University of California, San Francisco (UCSF) Institutional Review Board for all other datasets. All datasets were de-identified, and informed consent was obtained from all patients through these institutions. All methods were performed in accordance with relevant guidelines and regulations. 

\subsection{Evaluation}

\subsubsection{Quantitative Metrics}

We report standard generative metrics: Fréchet Inception Distance (FID) \cite{heusel_gans_2018} measuring how close the distribution of generated images is to real images, Multi-Scale Structural Similarity (MS-SSIM) \cite{wang_multiscale_2003} evaluating preservation of anatomical details, and Learned Perceptual Image Patch Similarity (LPIPS) \cite{zhang_unreasonable_2018} assessing how visually similar two images appear to a human observer. FID  was computed using Med3D network \cite{chen_med3d_2019} features for medical relevance. MS-SSIM and LPIPS, calculated separately inside and outside the tumor region to assess pathology preservation and anatomical diversity. 

To quantify sample diversity from repeated sampling of the same conditioning input, we computed pixel-wise mean and standard deviation across samples to assess spatial variation in generated outputs, as well as pairwise similarity metrics across all unique sample pairs. Cosine similarity was calculated by flattening each 3D volume into a 1D vector and computing the normalized dot product between sample pairs, where lower values indicate greater diversity. KL divergence was computed by binning intensity values into 50-bin histograms to form probability distributions for each sample, followed by calculating the Kullback–Leibler divergence between all pairs using the formula: $\text{KL}(P \parallel Q) = \sum_{x} P(x) \log \frac{P(x)}{Q(x)}$. Higher KL values indicate more diverse intensity distributions. For $N$ samples, this resulted in $N(N-1)/2$ pairwise comparisons, and the mean across all pairs was reported as the final diversity metric.

\subsubsection{Clinical Relevance}

We evaluate downstream utility via whole tumor segmentation using nnU-Net \cite{isensee_nnu-net_2021}. The baseline uses 1251 BraTS 2023 cases + 100 BraTS 2024 Additional cases. Synthetic augmentation experiments add 10, 50, or 100 generated images conditioned on real pathology extracted from the same 100 BraTS 2024 cases. This training set selection ensures the performance variation across models are introduced by the synthetic POWDR images, instead of additional pathologies in the training set. nnU-Net is trained for 1000 epochs with default parameters, and Dice scores are computed on 171 unseen test cases from BraTS 2024.  

Additionally, we assess anatomical plausibility by comparing brain tissue volumes (cerebral spinal fluid: CSF, gray matter: GM, and white matter: WM) between real and synthetic images using FSL FAST segmentation \cite{zhang_segmentation_2001}. We analyze 100 synthetic and 100 real cases, reporting volume statistics. 

\section{Results}

\subsection{Brain MRI}

\subsubsection{Qualitative Results}

\figureref{fig:fig2} illustrates three-plane views of the conditioning pathology image, the sampled synthetic image, and the original image where the conditioning pathology is extracted. The synthetic image differs substantially from the target image while preserving the pathology region. Brain structures are clearly visible, and tumors are seamlessly integrated without sharp edges or artifacts, indicating realistic anatomical synthesis. We also trained a conditional latent diffusion outpainting model (cLDM) \cite{rombach_high-resolution_2022} as a baseline for comparison. However, as shown in \figureref{fig:fig2}D, it failed to generate anatomically plausible images and was therefore excluded from the following quantitative analysis.

 \begin{figure}[htbp]
\floatconts
  {fig:fig2}
  {\caption{Qualitative example of brain MRI outpainting. Shown are A) the input condition (masked tumor), B) the synthetic image generated by POWDR, and C) the original image from which the tumor was extracted. Highlighted boxes indicate tumor regions preserved during synthesis. D) Synthetic image generated using a conditional Latent Diffusion outpainting model failed to converge. }}
  {\includegraphics[width=0.5\linewidth]{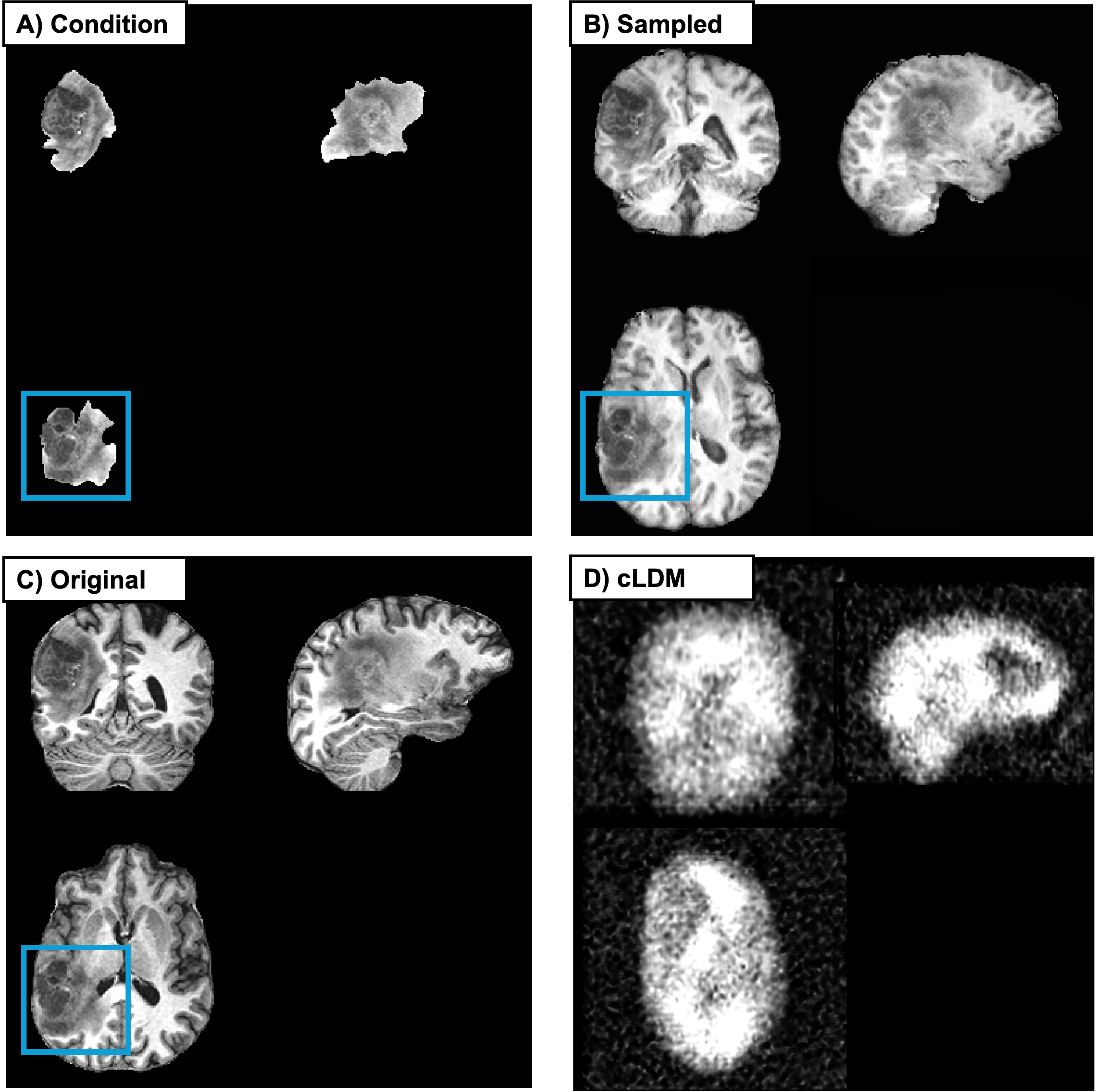}}
\end{figure}

\subsubsection{Quantitative Results}

We evaluate similarity to real images using FID, SSIM, and MSE (\tableref{tab:tab1}). Low FID scores of 0.0042 between generated and original cases indicates that synthesized images closely match the distribution of real data. For reference, the baseline FID score computed between two sets of real images is 0.0005. LPIPS analysis reveals high similarity within tumor regions (LPIPS-in 0.0030), confirming pathology preservation, and greater dissimilarity outside tumors (LPIPS-out 0.1484), suggesting successful augmentation of non-pathological areas. High MS-SSIM of 0.9998 inside tumor, 0.8195 outside of tumor further supports structural consistency within tumor region and diversity yet realism outside of tumor region. 

\begin{table}[htbp]
\floatconts
{tab:tab1}%
{\caption{Quantitative evaluation of image realism and diversity. Metrics include FID, MS-SSIM, and LPIPS, computed inside and outside the tumor region. Downward arrows indicate lower values are better; upward arrows indicate higher values are better; a dash (–) denotes mid-to-high values are optimal.}}%
    {\begin{tabular}{lll}
    \bfseries Metrics & \bfseries Mean (Std) - Real vs. Synthetic  & \bfseries Baseline - Real vs. Real\\
    FID $\downarrow$ & 0.0042 & 0.0005\\
    MS SSIM - inside tumor $\uparrow$ & 0.9998 (0.0003)& -\\
    MS SSIM - outside tumor - & 0.8195 (0.0517)&-\\
    LPIPS - inside tumor $\downarrow$ & 0.0030 (0.0017)&-\\
    LPIPS - outside tumor $\uparrow$ & 0.1484 (0.0311)&-\\
    \end{tabular}}
     
\end{table}

To assess diversity, we perform repeated sampling from the same conditioning input. \figureref{fig:fig3}A shows that when trained with masked tumor conditioning, generated samples exhibit small variation—an issue commonly observed in conditioned diffusion models. The mean sample across 20 repetitions appears nearly identical, and the standard deviation is low, indicating potential overfitting. To address this, we introduce random connected masks during training. \figureref{fig:fig3}B shows a substantial increase in standard deviation across repeated samples in regions outside the pathology, while remaining minimal within the conditioned pathology area, when trained with random connected masked region conditioning. The two example outputs further illustrate markedly variability in anatomical structures, especially in anatomical structures such as CSF cavities and brain boundaries. \figureref{fig:fig3}C further proves that strategy B significantly improves diversity: Cosine similarity between pairwise samples decreases from 0.9947 ± 0.00003 to 0.9580 ± 0.0066. KL divergence increases from 0.00026 to 0.01494. 

\begin{figure}[htbp]
\floatconts
  {fig:fig3}
  {\caption{Sampling diversity under different training strategies. A–B) Mean, standard deviation, and example outputs from 20-case repeated sampling using A) masked tumor conditioning and B) random connected mask conditioning. C) Similarity metrics cosine similarity and KL divergence quantify diversity improvements.}}
  {\includegraphics[width=0.6\linewidth]{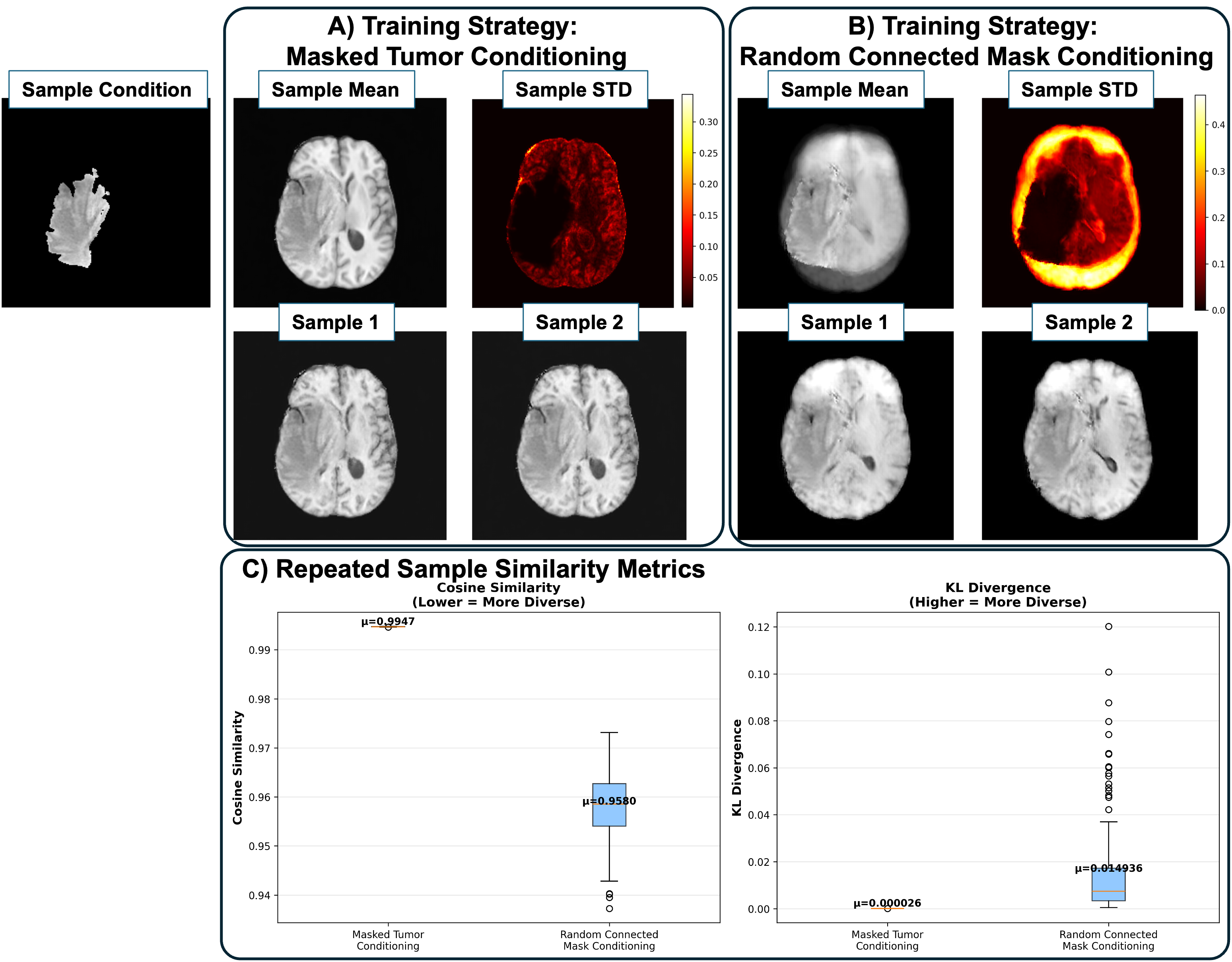}}
\end{figure}

This diversity enhancement is critical for downstream tasks. With tumor-mask conditioning (Strategy A), the maximum augmentation equals the number of unique conditions $N$, doubling the dataset at best. In contrast, random-mask training (Strategy B) enables generating $nN$ non-redundant samples, where $n$ is user-defined, which can be optimized to balance flexible augmentation without overwhelming the real data in downstream tasks. 

\subsubsection{Clinical Relevance}

We assess utility for tumor segmentation using nnU-Net. To isolate the effect of POWDR as data augmentation, synthetic images are generated from the same 100 BraTS 2024 cases used in segmentation training. Dice scores improve from 0.6992 ± 0.3009 (baseline) to 0.7015 ± 0.2955 with 10 synthetic cases, plateauing at 0.7137 ± 0.2868 for 50 cases and 0.7135 ± 0.2866 for 100 cases (\figureref{fig:fig4}). This suggests that moderate augmentation yields optimal gains.

\begin{figure}[htbp]
\floatconts
  {fig:fig4}
  {\caption{Impact of synthetic augmentation on tumor segmentation. Dice scores for nnU-Net segmentation under varying numbers of synthetic cases (10, 50, 100) compared to baseline. }}
  {\includegraphics[width=0.5\linewidth]{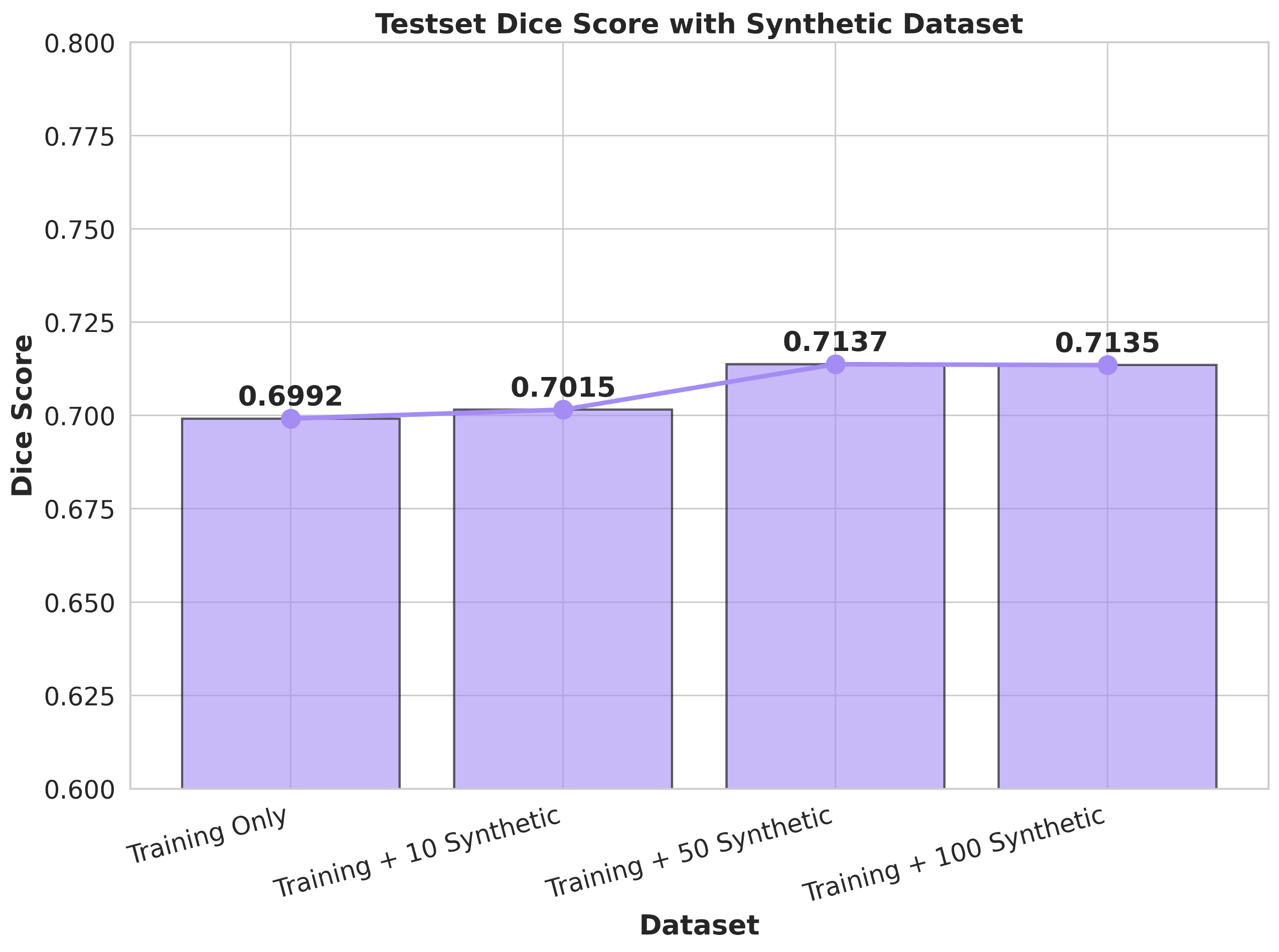}}
\end{figure}

For anatomical credibility, we perform brain tissue segmentation using FSL FAST algorithm. As shown in \figureref{fig:fig5}, volumes of CSF (Synthetic: 323.428 ± 32.001 mL, Real: 325.397 ± 49.228 mL) and GM (Synthetic: 534.475 ± 29.312 mL, Real: 524.283 ± 66.766 mL) show no significant difference from real images. WM volume is slightly lower in synthetic images, likely due to tumor presence affecting segmentation. 

 \begin{figure}[htbp]
\floatconts
  {fig:fig5}
  {\caption{Comparison of brain tissue volumes between real and synthetic images. Volumes of CSF, gray matter, and white matter segmented using FSL FAST for 100 synthetic and 100 real cases.}}
  {\includegraphics[width=0.7\linewidth]{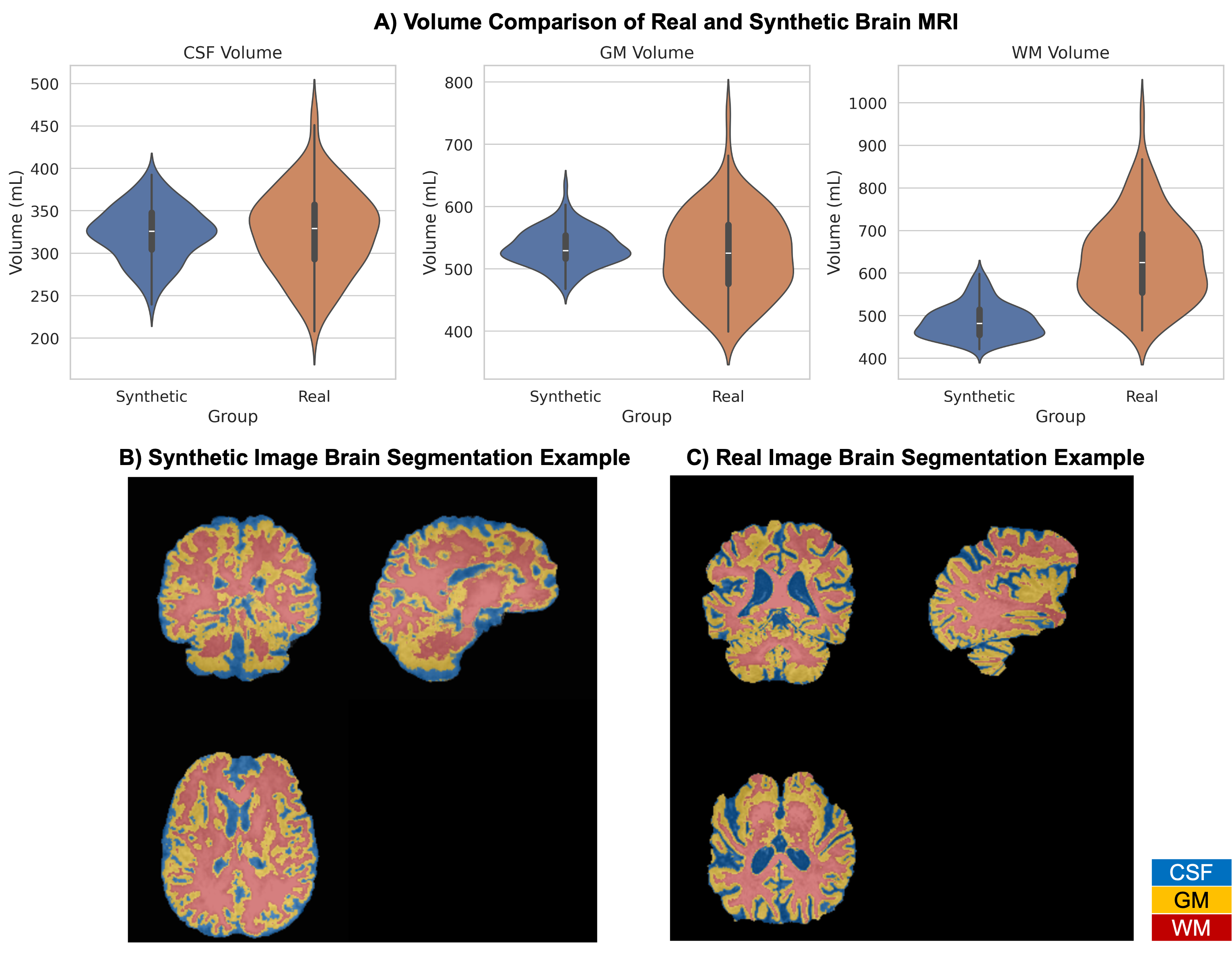}}
\end{figure}

\subsubsection{Knee MRI}

\figureref{fig:fig6} shows qualitative results for 3D knee MRI outpainting. Despite larger matrix size and longer training time, the model successfully preserves meniscus tears while generating realistic surrounding anatomy. This demonstrates the model is anatomy agnostic and highlights the method’s potential for pathology augmentation in low-prevalence conditions. We note that the contrast of the sampled images differs noticeably from the original images, likely due to variations in MR image contrast and scanner vendors present in the diffusion model’s training set.  

 \begin{figure}[htbp]
\floatconts
  {fig:fig6}
  {\caption{Qualitative results for knee MRI outpainting. The model preserves meniscus tears while generating realistic surrounding anatomy, demonstrating the framework is tissue agnostic.}}
  {\includegraphics[width=0.9\linewidth]{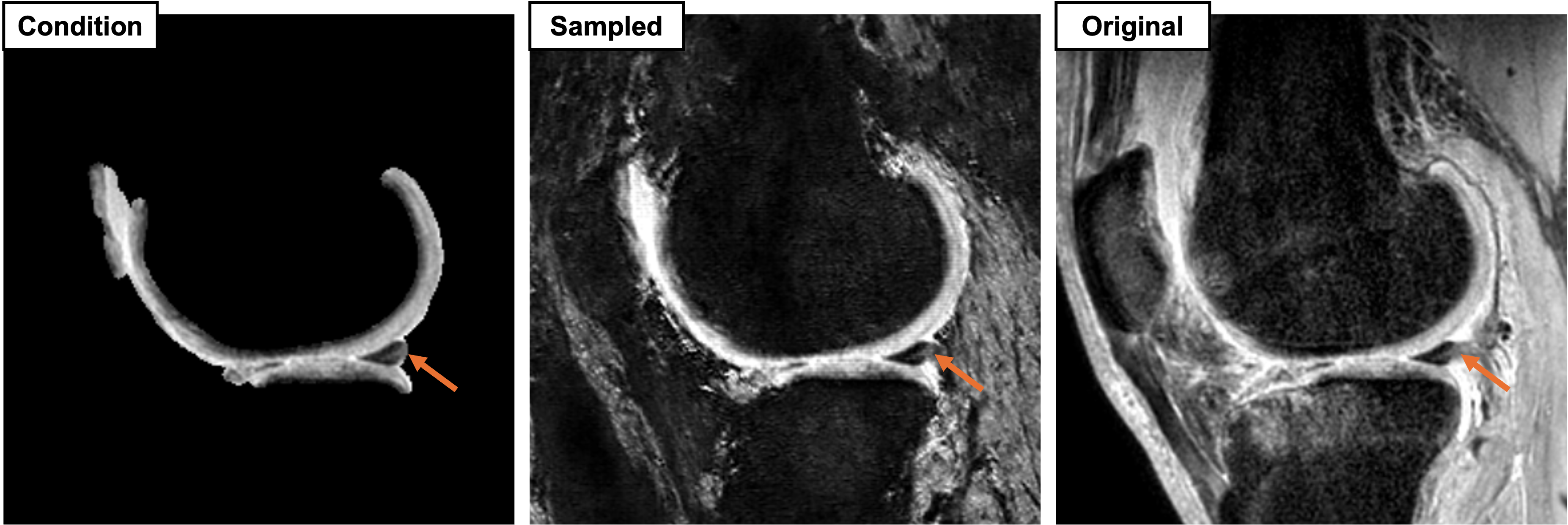}}
\end{figure}

\vspace{3cm}
\section{Conclusion}

In this work, we introduced POWDR, a pathology-preserving outpainting framework for 3D MRI based on a conditioned wavelet diffusion model. To our knowledge, this is among the first approaches to tackle outpainting in MRI while explicitly preserving real pathological regions. We conducted a comprehensive evaluation, including conventional generative metrics (FID, SSIM, LPIPS), diversity analysis through repeated sampling, and clinically relevant assessments such as tumor segmentation performance and brain tissue volume comparisons. Furthermore, we demonstrated the tissue-agnostic nature of our method by successfully applying it to knee MRI, highlighting its potential for broader anatomical domains. These results suggest that POWDR can serve as a practical solution for addressing data scarcity and class imbalance in 3D medical imaging, with wide-ranging applications in segmentation, classification, localization, and vision–language tasks.


%

\clearpage  

\bibliography{Diffusion}

%
%
%
%

\end{document}